\documentclass[nonacm]{acmart}
\usepackage{verbatim}
\usepackage{graphicx}
\usepackage{subcaption}
\usepackage{color}
\usepackage{csquotes}
\usepackage{xparse}
\usepackage{xcolor}
\usepackage{multirow}
\usepackage{tikz}
\usetikzlibrary{positioning,shapes.geometric}
\usepackage{float}
\begin{document}

\title{Building Trust Through Voice: How Vocal Tone Impacts User Perception of Attractiveness of Voice Assistants}

\author{Sabid Bin Habib Pias}
\email{sabhabib@iu.edu}
\authornotemark[1]
\affiliation{%
  \institution{Indiana University Bloomington}
  \country{USA}
}

\author{Alicia Freel}
\affiliation{%
  \institution{Indiana University Bloomington}
  \country{USA}
}
\email{anfreel@iu.edu}

\author{Ran Huang}
\affiliation{%
  \institution{Indiana University Bloomington}
  \country{USA}
}
\email{huangran@iu.edu}

\author{Donald Williamson}
\affiliation{%
  \institution{Ohio State University}
  \country{USA}
}
\email{williamson.413@osu.edu}

\author{Minjeong Kim}
\affiliation{%
  \institution{Indiana University Bloomington}
  \country{USA}
}
\email{kim2017@indiana.edu}

\author{Apu Kapadia}
\affiliation{%
  \institution{Indiana University Bloomington}
  \country{USA}
}
\email{kapadia@indiana.edu}

\renewcommand{\shortauthors}{Pias et al.}

\begin{abstract}
  Voice Assistants (VAs) are popular for simple tasks, but users are often hesitant to use them for complex activities like online shopping. We explored whether the vocal characteristics like the VA's vocal tone, can make VAs perceived as more attractive and trustworthy to users for complex tasks. Our findings show that the tone of the VA voice significantly impacts its perceived attractiveness and trustworthiness. Participants in our experiment were more likely to be attracted to VAs with positive or neutral tones and ultimately trusted the VAs they found more attractive. We conclude that VA's perceived trustworthiness can be enhanced through thoughtful voice design, incorporating a variety of vocal tones.
\end{abstract}





\begin{teaserfigure}
  \centering
    \includegraphics[width=0.9\linewidth]{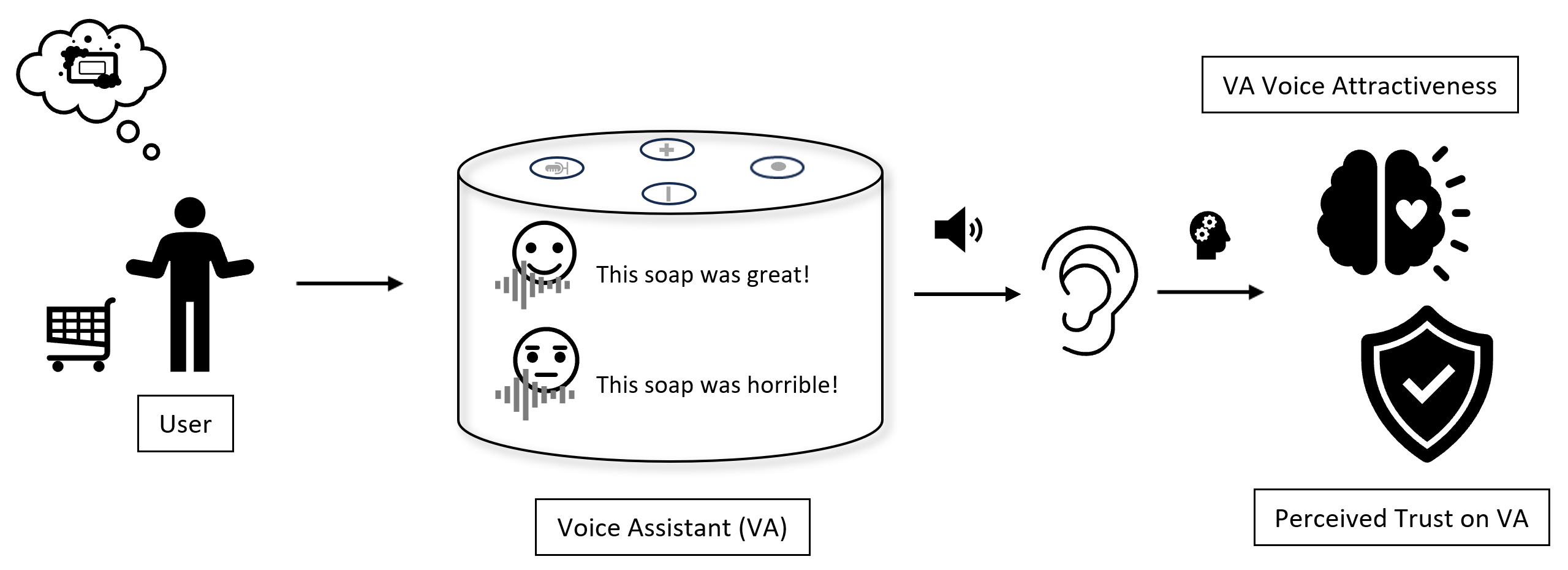}
    \caption{An illustration of the experiment depicts a user interacting with a voice assistant (VA) featuring diverse vocal tone variations. We hypothesize that the tones of the VA voice influence VA's perceived attractiveness, and the perceived attractiveness subsequently impacts the perceived trustworthiness of the VAs.}
    \label{fig:voice-attract-emotion-overview}
\end{teaserfigure}

\maketitle

\section{Introduction and Related Work}

Rapid advancements in generative AI and natural language processing (NLP) have significantly expanded the capabilities of voice assistants (VAs)~\cite{10.1145/3580305.3599572, dighe2024leveraging}, enabling them to perform complex tasks such as engaging in in-depth conversations or providing e-shopping recommendations~\cite{de2020reducing, alexashopping}. Despite these advancements, user adoption for intricate tasks remains limited due to a lack of trust in the VA~\cite{novozhilova_looking_2024, vaownernerve}. This mistrust stems from various concerns, including the skepticism that VAs might make mistakes~\cite{campbell_moderating_2001, muthukumaran_optimizing_2020, hong_what_2020} or interactions with VAs may not be as engaging as those with humans~\cite{jain2023impact}. Therefore, increasing user trust and engagement is crucial for expanding VA adoption in more intricate tasks.

Research suggests that enhancing the perceived physical attractiveness of VAs can serve as a strategy to improve user engagement and trust in VAs~\cite{rheu2021systematic}. Given the significant implications of the disembodied nature of VAs, such as their interactions with individuals with visual impairments~\cite{mina2023leveraging}, we aim to investigate alternative methods to enhance VA attractiveness. Specifically, we investigate whether acoustic attractiveness can similarly boost user trust and engagement in performing complex tasks with VAs. Introducing human-like qualities in the VAs can increase user attraction to VAs by reducing psychological distance~\cite{beattie2020bot, 10.1145/3479515}. We investigate the vocal tone of VAs as an anthropomorphic quality, as studies demonstrated the effectiveness of human vocal tones in enhancing attractiveness in interpersonal relationships~\cite{zuckerman1989sounds}. Similar effects of vocal attributes have also been observed in human-robot interactions. For example, users generally prefer robots with warm and engaging voices over those with a robotic or monotone quality~\cite{nass2001effects}. Furthermore, urgency in the vocal tone of VAs has been effective in emergency drill scenarios~\cite{kim_urgency_2023}. Therefore, diversifying the tone of voice in VAs has the potential to increase their perceived attractiveness for users and subsequently enhance the perceived trustworthiness of the VAs. However, recent VAs offer limited voice customization. Alexa provides one male and one female voice per accent~\cite{AmazonAlexa}. Google Assistant has 12 different voices with varied accents~\cite{googlecolor}, and Siri offers a few male and female voices with different accents~\cite{sirivoice}. Notably, these options lack diversity in vocal tone. 

Studies show that users prefer voices that sound engaging and relatable~\cite{zhong2022user}. Therefore, customizing VA voices to match user preferences may make users feel more comfortable and trusting of VAs, especially for important tasks such as online shopping, banking, and bill payments. Increasing comfort can help more people, including the elderly and those with disabilities, use voice assistants for complex tasks if they can trust VAs more with these specific tasks~\cite{10.1145/3373759, 10.1145/3441852.3471218, vieira2022impact}. However, the effects of varying vocal tones on the perceived attractiveness of VAs and their influence on user trust remain largely unexplored. This gap in research underscores the need for further investigation into how specific vocal attributes can enhance the attractiveness and trustworthiness of VAs in consumer contexts. To address this gap, our aim is to investigate how modifying the vocal tone of VAs affects the perceived attractiveness and trustworthiness of VAs. Specifically, we seek to answer the following question:

\textit{\textbf{RQ1:}} How does the perceived tone of a voice assistant's voice affect its perceived attractiveness and subsequently affect people's trust in the voice assistant?

We aimed to explore how positive, negative, and neutral tones of voices by VAs enhance their perceived attractiveness in users, and consequently, whether the participants' trust is affected by the perceived attractiveness. 

It is also important to consider the ethical implications of deploying diverse vocal tones to enhance VA's attractiveness and trustworthiness, for example, where trust may be misplaced or unwarranted. Our ultimate goal is to assist users in making better decisions while increasing their engagement with VAs. Therefore, it is imperative that the VAs are indeed trustworthy; they should provide factual information, perform to user expectations, and communicate with users to increase user comfort and trust~\cite{ma2020challenges}, along with improving the attractiveness of the voice of the VAs.

This poster presents additional analysis beyond our previous study~\cite{pias2024impact}, which investigated whether users are persuaded by the vocal tones of voice assistants (VAs) and if this persuasiveness influences their purchase intentions. Building on the same study, we examine the interaction between VA and participants from another perspective: whether the tone of VA voice enhances acoustic attractiveness, and whether this increased attractiveness leads to greater trust in the VAs.
\section{Methodology}

We generated voice stimuli with varying tones, age groups, and genders and  validated the voice stimuli in an initial study, and then measured user behavior toward varying VA voices in a subsequent study. All studies were verified and approved by the ethics review board (IRB) of Indiana University.

We followed the stimuli generation process by Waller et al.~\cite{skoog2015can} to generate voices that varied by tone, age group, and gender. Male and female voices categorized by age groups (younger adults 20--30 years, middle-aged adults 40--50 years, and older adults 60--70 years) were selected, and positive (happy, excited, cheering), neutral (default/flat tone) or negative (sad, frustrated) tones were applied via Microsoft Studio\footnote{\url{https://speech.microsoft.com/}}. The audio was processed using Audacity\footnote{\url{https://www.audacityteam.org/}} to standardize the intensity~\cite{skoog2015can}. We chose the most popular~\cite{statistamostpurchase, bluecartmostpurchase} gender-neutral products and their reviews from Amazon\footnote{\url{https://www.amazon.com/}}. Positive product reviews were generated in either a positive or neutral tone, while negative reviews were generated in either a negative or neutral tone. Each review contained 26 to 33 words per Waller et al. ~\cite{skoog2015can}, resulting in audio clips between 12 and 16 seconds. 

We conducted a stimulus validation study to identify the perceived tone and age group of the generated voices (N = 78)~\cite{baird2017perception, baird2018perception}. We removed older adult voices from the stimulus set, as pilot study participants could not detect the age of older adult voices properly. In the final stimulus validation survey,
each participant was randomly assigned to either a male or female voice to eliminate gender bias. Participants rated the tone of the voice~\cite{amon2020influencing} and estimated the age group (younger adult or middle-aged adult). The average ratings were calculated, categorizing the voices by tone and age group.
 
The main study utilized the validated stimuli to investigate user perceptions of the varying vocal characteristics of a VA using the Prolific\footnote{\url{https://app.prolific.co/}}. We asked the participants (N = 335) to imagine themselves checking product reviews using a voice assistant. We conducted the study with a factorial design of 2x2x2x2 between subjects, examining the effects of the review valence (positive vs. negative), gender of the voice assistant (female vs. male), age (younger vs. middle-aged) and tone (positive\textbackslash negative vs. neutral). Each participant was presented with one specific voice type, with the between-factor design used to prevent carry-over effects. Participants listened to these reviews individually, which were 26 to 35 words long and lasted between 12 and 17 seconds. We measured participants' perceived attractiveness of the voice and perceived trustworthiness of the VA (Appendix table~\ref{tab:ques-voice}).

Next, we presented the participants with two qualitative open-ended questions:

\begin{itemize}
    \item \textit{What do you like/dislike about the voice you heard during the survey? Please elaborate.}
    \item \textit{If you could create a voice for the voice assistant, what would that voice sound like? Please elaborate.}
\end{itemize}
 
Based on a priori G*Power analysis~\cite{faul2007g} we used the responses of 335 participants for further analyses, where more than 94\% participants had experience using a voice assistant. All participants (stimulus included) were paid according to the minimum wage recommendation in the study location~\cite{silberman2018responsible}. We used Cronbach's alpha~\cite{cronbach1951coefficient} to assess internal consistency among the dependent variables. Post-regression, we calculated pairwise comparisons using estimated marginal means to interpret significant interaction effects. We conducted an inductive thematic analysis on open-ended questions about participants' likability for the VA voice. Using Delve\footnote{\url{https://delvetool.com/}}, we coded responses and identified themes. We developed and used a code book to code all responses and derive the main themes.

Research has shown that vocal attributes influence listeners' perceived attractiveness of the speaker in interpersonal communication~\cite{zuckerman1989sounds}. Moreover, physical attractiveness has been shown to enhance trust in human interactions~\cite{zhao2015face}.  We aimed to explore the relationship between vocal characteristics, perceived attractiveness, and trustworthiness of the speaker in the context of voice assistants. We selected vocal tone as a vocal characteristic and investigated how the perceived tone of a VA's voice affects the VA attractiveness and how this, in turn, impacts users' trust in the VA's recommendations. Based on our research questions, we propose the following hypothesis.

\textit{\textbf{H1:}} The \textit{positive} tone of a voice assistant's voice enhances participants' perceived attractiveness of the VAs more compared to the \textit{neutral} or \textit{negative} tones and subsequently, the enhanced attractiveness increases participants' trustworthiness of the VA.

\section{Results}
\subsection{Effect of Vocal Tones}
We employed the causal mediation analysis framework~\cite{baron_moderatormediator_1986, tingley_mediation_2014} using R package `lavaan'~\cite{rosseel_lavaan_2012} to estimate the direct effect of the perceived tone of the VA voice on the perceived trustworthiness of the VAs and the indirect effect through the perceived VA voice attractiveness. We included the perceived age and gender of the voice in the mediation model to intercept possible indirect or direct effects of voice age and gender on perceived VA trustworthiness.  We used the comparative fit index (CFI)~\cite{bentler_comparative_1990} and significant model test statistics to select an optimal mediation model. In addition, the data satisfied the assumptions of homoscedasticity and normality of residuals. Thus, we performed linear regression to evaluate the relationship between the attractiveness of the VA voice and the VA voice tones.  

\begin{figure}[ht!]
\centering
\begin{tikzpicture}[>=stealth, node distance=1cm]
\tikzstyle{var} = [rectangle, draw, text centered, minimum height=15mm, minimum width=15mm]
\tikzstyle{latent} = [diamond, draw, text centered, minimum size=2em, fill=gray!50]
\tikzstyle{edge} = [->, thick]

\node[var] (emt)[] {Tone};
\node[var] (rv) [below=of emt] {Review Valence};



\node[var] (va) [right=of emt, xshift=0.2cm, yshift=2cm] {VA Voice Attractiveness};
\node[var] (vt) [right=of rv, xshift=2cm, yshift=1cm] {VA Trustwothiness};

\draw[edge] (rv) -- (vt) node[near end, above] {$-0.47^{**}$};
\draw[edge] (emt) -- (vt) node[near start, above] {$-0.20$};

\draw[edge] (emt) -- (va) node[midway, above] {$0.54^{***}$};

\draw[edge] (va) -- (vt) node[midway, right] {$0.32^{***}$};

\end{tikzpicture}
\caption{Path model diagram for the mediation of voice attractiveness in the relationship between VA trustworthiness and the tone, and review valence of the VA voice. (* = p < 0.05, ** = p < 0.01, *** = p < 0.001). Baselines for emotion=neutral}
\label{fig:path-diagram}
\end{figure}
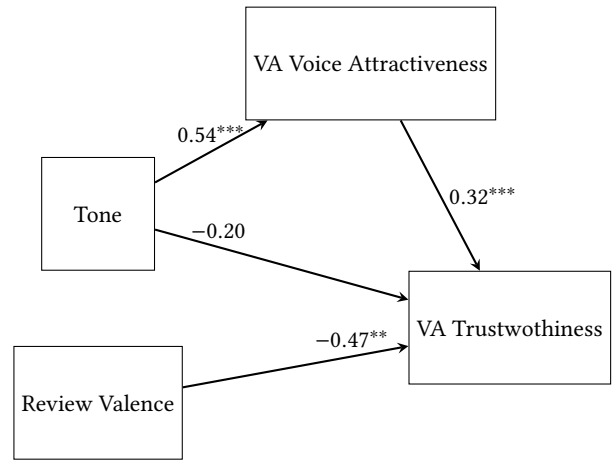

\begin{figure}[ht!]
  \centering
  \begin{minipage}[b]{0.45\textwidth}
    \centering
    \includegraphics[width=0.95\textwidth]{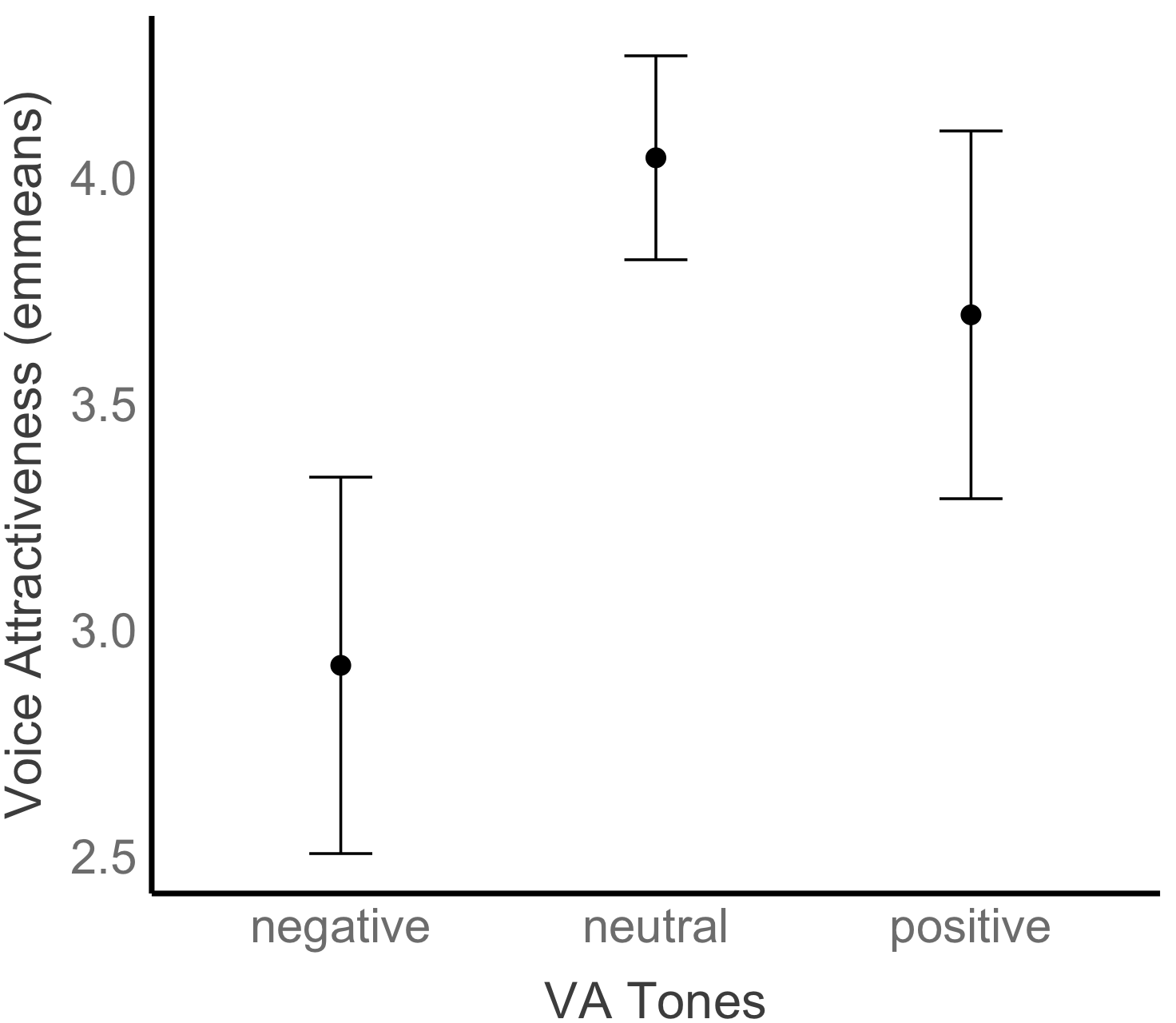}
    \caption{Differences in the perceived attractiveness of voice in terms of vocal tones (estimated marginal means)}
    \label{fig:voice-attract-emotion}
  \end{minipage}
  \hfill
  \begin{minipage}[b]{0.45\textwidth}
    \centering
  \begin{tabular}{llll}
        \hline
        & Estimate & Std. Error\\
      \hline
       Tone & $0.18^{***}$ & 0.04   \\
       Gender & $0.1$ & 0.06   \\
       Age & $-0.09$ & 0.06   \\
     \hline
  \end{tabular}

\caption{Indirect effect of VA tone, age, and gender in the proposed mediation model (* = p < 0.05, ** = p < 0.01, *** = p < 0.001)}
\label{tab:effects-mediation}
  \end{minipage}
\end{figure}

Table~\ref{tab:effects-mediation} shows a significant indirect effect of VA tone on trustworthiness, mediated by the attractiveness of the VA's voice. VA tone directly affected the attractiveness of the voice, which then affected the trust of the participants (Figure~\ref{fig:path-diagram}). A linear regression analysis indicated that positive tones (medium effect) and neutral tones (medium effect) were perceived as more attractive than negative tones (Figure~\ref{fig:voice-attract-emotion}). These results partially support hypothesis \textit{H1}. Although positive tones significantly enhanced attractiveness compared to negative tones, there was no significant difference in perceived attractiveness between positive and neutral tones.

\emph{In summary, positive and neutral VA tones were perceived as more attractive than negative tones, and the perceived attractiveness of the VA voice significantly influenced perceived trustworthiness in the VAs. Positive and neutral tones similarly affected VA's voice attractiveness and, consequently, VA's trustworthiness. However, for negative reviews, neutral tones were perceived as more attractive than negative tones, leading participants to trust negative reviews more when presented with neutral tones compared to negative VA tones.}

Interestingly, we did not find any significant indirect or direct effect of VA voice gender and age on perceived VA trustworthiness. 

\subsection{Reasons for Voice Preference}

We analyzed participants' written responses to understand their preferences for voice tones and how those preferences influenced trust. 

Some participants found positive-toned VAs more believable and genuine. For example, P36 stated that the positive toned VA ``\textit{felt and sounded authentic and spoke pretty naturally}''. The authenticity may have been contributed by the comfortable atmosphere created by the VA that ``\textit{sounded friendly and positive}'' (P97).

In addition, some participants preferred neutral voices. They perceived the VA's calmness as a sign of unbiased recommendations as the VA ``\textit{did not give off any type of extreme emotion}'' (P87). Moreover, a few participants felt that the neutral-toned VA was more fair as it ``\textit{was giving more factual statements, rather than opinions.}'' (P15).

Some participants did not like the negative tone of VA voices because they found the strong emotion distracting and overwhelming. P19 stated, ``\textit{The voice with negative tone seemed pessimistic towards the majority of things}.'' Some participants found it harder to concentrate on the content as ``\textit{the voice with a negative tone sounded uninterested or almost sad}'' (P146). This group of participants was more interested in the information than in how it was delivered.

\section{Discussion and Future Work}

In our study, we observed that participants found neutral and positive tones more attractive than negative tones, indirectly influencing their trust in VAs. The similar attractiveness of neutral and positive tones is interesting because positive attitudes and attractiveness are often correlated~\cite{reinhard2006explicit}, and positive emotions have been shown to be more attractive in various contexts~\cite{beattie2020bot}. This result is similar to our previous work that found that a group of users find calm and positive tones to be more persuasive in online shopping scenarios~\cite{pias2024impact}. The written responses of the participants indicated that some people preferred positive tones because a bit of entertainment makes complex tasks more enjoyable. Conversely, a group of participants found the robotic tones to be more appealing, as these tones were factual and allowed them to focus on the recommendations without being distracted. In contrast, some participants expressed that the negative tone of the VAs sounded uninterested and overwhelming, hindering their ability to concentrate on the information. Thus, negative tones were generally not favored by participants. This diverse result reinforces the notion that attractiveness is not always objective~\cite{nestor2010subjective}; what one person finds attractive, another may not.

In our study, we have also found that when participants were more attracted to VAs, their perceived trustworthiness of the VAs increased significantly. This finding is consistent with prior research that users often associate attractive interfaces with better usability~\cite{linghammar2007usability}, consequently leading to higher user satisfaction~\cite{lindgaard2007aesthetics}. Thus, an appealing interface promotes long-term use and loyalty, which in turn increases user trustworthiness in the interface. Similarly, an attractive VA can create a positive first impression, setting the tone for the entire user experience. Attractive designs can evoke positive emotions, making users feel more comfortable and confident while using the interface. Research has shown that people remember an event more when that event can invoke an emotional response in them~\cite{laird1982remembering}. Therefore, when positive or neutral tones enhanced the attractiveness of VAs, they subsequently invoked trustworthiness. A friendly VA voice can also calm people down, making them feel more comfortable using it for even personal tasks. 

Making voice assistants more attractive and thus increasing trustworthiness can help reach a broader audience of people who can utilize VAs as a tool, including those who may find traditional interfaces challenging (such as the elderly or visually impaired)~\cite{khan2021insight}. In addition to features such as jokes or friendly responses~\cite{moussawi2021effect}, the diverse vocal tones of VAs can make the experience fun and engaging. Making VAs more attractive through diverse vocal tones also helps reduce the barrier to use VAs, as some people find machines with anthropomorphic attributes easier to interact with~\cite{pelau2021makes}. A pleasant and engaging voice assistant can create a stronger emotional connection with users~\cite{flavian2021impacts}, making them more likely to enjoy using the technology and subsequently build user trust over time.

Future VA designs should utilize the effects of diverse VA tones to increase user engagement, considering the positive effect of positive and neutral VA tones on participants' perceived attractiveness and trust in the VAs. However, the personality traits of the users can influence their perception of attractiveness and trust~\cite{bartosik2021you}. Moreover, the user acceptance of new information or features can be dictated by their personality~\cite{pias2024drawback}.  Therefore, VA designs should account for diverse personality-based preferences to ensure that VA voices are appealing and acceptable to a wide range of users. 

It is also important to remember that diverse vocal tones are only one element within a broader ethical framework. For VAs to establish genuine trust, they must be trustworthy and demonstrate consistent, accurate performance, and effective communication alongside an appealing vocal style. A VA's overall experience, including voice, performance, and communication methods, builds trust and credibility. Users are more likely to rely on a VA that consistently provides accurate information and helpful assistance in a clear and professional manner.
\section{Limitation}
Our study has some limitations. First, we only used male and female voices, as technology for generating gender-ambiguous voices~\cite{sutton_gender_2020} with specific tones and ages is still developing. Second, the online setting with computer voices might not capture real-world interactions with physical VAs or the user's natural shopping environment. 
\section{Conclusion}

We study how the vocal tone of voice assistants (VAs) affects their perceived attractiveness to users and, in turn, their trustworthiness.  
Our findings indicate that positive and neutral tones increase the perceived voice attractiveness, which consequently increases user trust in the VAs. Based on these results, we suggest that incorporating a greater diversity of vocal tones can be an effective strategy to make VAs more attractive and trustworthy to users. This approach can lower barriers to VA usage, increase user engagement, and improve communication effectiveness through enhanced trust. In addition, we emphasize that VA tones should be carefully designed to avoid misleading users: rather, the tone of VA voice should serve as a medium to foster healthy engagement with trustworthy VAs.

\begin{acks}
   This material is based upon work supported by the National Science Foundation under grants CNS-2207019 and Social Science Research Funding Program (SSRFP) in Indiana University. We would also like to thank Hannah Bolte and Elizabeth Ray for providing statistics consultation from the Indiana Statistical Council Center (ISCC). 
\end{acks}

\bibliographystyle{ACM-Reference-Format}
\bibliography{ref-auth1, references}

\clearpage
\appendix
\section{Appendix}
\begin{table}[ht]
  \begin{tabular}{p{3cm}p{3cm}p{1.5cm}}
        \hline
      Factor & Item & Chron. alpha\\
     \hline
      \multirow{3}{*}{\parbox{3cm}{\centering VA Voice\\Attractiveness~\cite{park2020effects}}} & The voice gave me a good feeling & \multirow{3}{*}{0.88}  \\
       & The voice is attractive &   \\
       & The voice caught my attention &    \\
       \hline
      \multirow{5}{*}{\parbox{3cm}{\centering VA Voice\\Trustworthiness~\cite{ohanian1990construction}}} 
      & The voice assistant is dependable & \multirow{5}{*}{0.91}  \\
      & The voice assistant is honest  & \\
      & The voice assistant is sincere  &  \\
       & The voice assistant is trustworthy &   \\
       & The voice assistant is reliable &   \\
       \hline
  \end{tabular}
  \caption{Question Items about Perceived Voice Attractiveness and Perceived VA Trustworthiness. Chon. alpha= Chronbach's alpha}
  \label{tab:ques-voice}
\end{table}

\end{document}